\documentclass[aps,prd,twocolumn,superscriptaddress]{revtex4}
\usepackage{epsfig,epsf}
\usepackage{amsmath}
\usepackage{amsthm}
\usepackage{amsfonts}
\usepackage{amssymb}
\usepackage{dsfont}
\usepackage{multirow}
\usepackage{appendix}
\usepackage{slashed}
\usepackage[active]{srcltx}
\usepackage{psfrag}

\begin{document}

\title{Decay modes of the scalar exotic meson $T_{bs;\overline{u}\overline{d}%
}^{-}$}
\date{\today}
\author{S.~S.~Agaev}
\affiliation{Institute for Physical Problems, Baku State University, Az--1148 Baku,
Azerbaijan}
\author{K.~Azizi}
\affiliation{Department of Physics, University of Tehran, North Karegar Ave., Tehran
14395-547, Iran}
\affiliation{Department of Physics, Do\v{g}u\c{s} University, Acibadem-Kadik\"{o}y, 34722
Istanbul, Turkey}
\author{H.~Sundu}
\affiliation{Department of Physics, Kocaeli University, 41380 Izmit, Turkey}

\begin{abstract}
We investigate the semileptonic decay of the scalar tetraquark $T_{bs;%
\overline{u}\overline{d}}^{-}$ to final state $T_{cs;\overline{u}\overline{d}%
}^{0} l \overline{\nu}_l$, which proceeds due to the weak transition $b \to
c l \overline{\nu}_l$. For these purposes, we calculate the spectroscopic
parameters of the final-state scalar tetraquark $T_{cs;\overline{u}\overline{%
d}}^{0}$. In calculations we use the QCD sum rule method by taking into
account the quark, gluon, and mixed condensates up to dimension 10. The mass
of the $T_{cs;\overline{u}\overline{d}}^{0}$ obtained in the present work $(
2878 \pm 128 )~\mathrm{MeV}$ indicates that it is unstable against the
strong interactions, and can decay to the mesons $D^{0}\overline{K}^{0}$ and
$D^{+}K^{-}$. Partial widths of these $S$-wave modes as well as the full
width of the tetraquark $T_{cs;\overline{u}\overline{d}}^{0}$ are found by
means of the QCD light-cone sum rule method and technical tools of the
soft-meson approximation. The partial widths of the main semileptonic
processes $T_{bs;\overline{u}\overline{d}}^{-} \to T_{cs;\overline{u}%
\overline{d}}^{0}l\overline{\nu}_l$, $l=e, \mu$, and $\tau$ are computed by
employing the weak form factors $G_{1}(q^2)$ and $G_{2}(q^2)$, which are
extracted from the QCD three-point sum rules. We also trace back the weak
transformations of the stable tetraquark $T_{bb;\overline{u}\overline{d}%
}^{-} $ to conventional mesons. The obtained results for the full width $%
\Gamma_{\mathrm{full}}=(3.28\pm 0.60) \times 10^{-10}~\mathrm{MeV} $ and
mean lifetime $\tau=2.01_{-0.31}^{+0.44}~\mathrm{ps}$ of $T_{bs;\overline{u}%
\overline{d}}^{-}$, as well as predictions for decay channels of the
tetraquark $T_{bb;\overline{u}\overline{d}}^{-}$ can be used in experimental
studies of these exotic states.
\end{abstract}

\maketitle

%%%%%%%%%%%%%%%%%%%%%%%%%%%%%%%%%%%%%%%%%%%%%%%%%%%%%%%%%%%%%%%%%%

\section{Introduction}

\label{sec:Int}
%%%%%%%%%%%%%%%%%%%%%%%%%%%%%%%%%%%%%%%%%%%%%%%%%%%%%%%%%%%%%%

Investigation of exotic mesons composed of four valence quarks, i.e.,
tetraquarks is one of the interesting and intriguing problems on agenda of
high energy physics. Experimental data collected by various collaborations
and achievements in their theoretical explanations made these states an
important part of hadron spectroscopy \cite%
{Chen:2016qju,Chen:2016spr,Esposito:2016noz,Ali:2017jda,Olsen:2017bmm}. But
the nonstandard mesons discovered till now and considered as candidates to
exotics are wide resonances which decay strongly to conventional mesons.
These circumstances obscure their four-quark bound-state nature and inspire
appearance of alternative dynamical models to account for observed effects.
Therefore, theoretical and experimental studies of 4-quark states, which are
stable against the strong interactions can be decisive for distinguishing
dynamical effects and genuine multiquark states from each another.

The problems of stability of 4-quark mesons were already addressed in the
original papers \cite{Ader:1981db,Lipkin:1986dw,Zouzou:1986qh}. The
principal conclusion made in these works was that, if a mass ratio $%
m_{Q}/m_{q}$ is large, then the heavy $Q$ and light $q$ quarks may
constitute stable $QQ\bar{q}\bar{q}$ compounds. The stabile nature of the
axial-vector tetraquark $T_{bb;\overline{u}\overline{d}}^{-}$ (briefly, $%
T_{bb}^{-}$) was predicted in Ref.\ \cite{Carlson:1987hh}, and confirmed by
recent investigations \cite{Navarra:2007yw,Karliner:2017qjm,Eichten:2017ffp}%
. The similar conclusions about the strong-interaction stability of the
tetraquarks $T_{bb;\overline{u}\overline{s}}^{-}$, and $T_{bb;\overline{d}%
\overline{s}}^{0}$ were drawn in Ref.\ \cite{Eichten:2017ffp} as well. The
spectroscopic parameters and semileptonic decays of the axial-vector
tetraquark $T_{bb;\overline{u}\overline{d}}^{-}$ were analyzed in our work
\cite{Agaev:2018khe}. Our result for the mass of the $T_{bb}^{-}$ state $%
(10035\pm 260)~\mathrm{MeV}$ is below the $B^{-}\overline{B}^{\ast 0}$ and $%
B^{-}\overline{B}^{0}\gamma $ thresholds, respectively, which means that it
is strong- and electromagnetic-interaction stable particle and can decay
only weakly. We evaluated the full width and mean lifetime of $T_{bb}^{-}$
using its semileptonic decay channel $T_{bb}^{-}\rightarrow Z_{bc}^{0}l\bar{%
\nu _{l}}$ (for simplicity, $Z_{bc}^{0}\equiv Z_{bc;\bar{u}\bar{d}}^{0}$).
The predictions $\Gamma =(7.17\pm 1.23)\times 10^{-8}~\mathrm{MeV}$ and $%
\tau =9.18_{-1.34}^{+1.90}~\mathrm{fs}$ obtained in Ref. \cite{Agaev:2018khe}
are useful for further experimental studies of this double-heavy exotic
meson.

Because the tetraquark $T_{bb}^{-}$ decays dominantly to the scalar state $%
Z_{bc}^{0}$, in Ref. \cite{Agaev:2018khe} we calculated also the
spectroscopic parameters of $Z_{bc}^{0}$. The mass of this state $(6660\pm
150)~\mathrm{MeV}$ is considerably below $7145~\mathrm{MeV}$ required for
strong decays to heavy mesons $B^{-}D^{+}$ and $\overline{B}^{0}D^{0}$. The
threshold for electromagnetic decays of $Z_{bc}^{0}$ exceeds $7600~\mathrm{%
MeV}$, and is also higher than its mass. The semileptonic and nonleptonic
weak decays of the tetraquark $Z_{bc}^{0}$ were explored in Ref.\ \cite%
{Sundu:2019feu}. The dominant weak decay modes of $Z_{bc}^{0}$ contain at
the final state the scalar tetraquark $T_{bs;\overline{u}\overline{d}}^{-},$
which has the mass $(5380\pm 170)~\mathrm{MeV}$, and is strong- and
electromagnetic-interaction stable particle.

The spectroscopic parameters and width of the axial-vector state $T_{bc}^{0}$
with the same quark content $bc\overline{u}\overline{d}$ were computed in
Ref.\ \cite{Agaev:2019kkz}. The central value of its mass $(7105\pm 155)~%
\mathrm{MeV}$ is lower than corresponding thresholds both for strong and
electromagnetic decays. Both the semileptonic and nonleptonic weak decays of
$T_{bc}^{0}$ create at the final state the scalar tetraquark $T_{cc;%
\overline{u}\overline{d}}^{+}$, which is strong-interactions unstable
particle and decays to conventional mesons $D^{+}D^{0}$ \cite{Agaev:2019qqn}.

The 4-quark compounds $bc\overline{u}\overline{d}$ were subjects of
interesting theoretical studies \cite%
{Karliner:2017qjm,Eichten:2017ffp,Feng:2013kea,Francis:2018jyb,Caramees:2018oue}%
. Thus, an analysis performed in Ref.\ \cite{Karliner:2017qjm} showed that $%
Z_{bc}^{0}$ lies below the threshold for $S$-wave decays to conventional
heavy mesons, whereas the authors of Ref. \cite{Eichten:2017ffp} predicted
the masses of the scalar and axial-vector $bc\overline{u}\overline{d}$
states above the $B^{-}D^{+}/\overline{B}^{0}D^{0}$ and $B^{\ast }D$
thresholds, respectively. Nevertheless, explorations conducted using the
Bethe-Salpeter method \cite{Feng:2013kea}, and recent lattice simulations
proved the strong-interaction stability of the axial-vector exotic meson $%
T_{bc}^{0}$ \cite{Francis:2018jyb}. An independent analysis of Ref.\ \cite%
{Caramees:2018oue} also confirmed the stability of the tetraquarks $bc%
\overline{u}\overline{d}$; it was demonstrated there, that both the scalar
and axial-vector states $bc\overline{u}\overline{d}$ are stable against the
strong interactions.

Summing up one sees, that $T_{bb}^{-}$ transforms due to chains of the
decays $T_{bb}^{-}\rightarrow Z_{bc}^{0}l\bar{\nu _{l}}\rightarrow T_{bs;%
\overline{u}\overline{d}}^{-}l\bar{\nu _{l}}\overline{l^{\prime }}\nu
_{l^{\prime }}$ and $T_{bb}^{-}\rightarrow Z_{bc}^{0}l\bar{\nu _{l}}%
\rightarrow T_{bs;\overline{u}\overline{d}}^{-}Pl\bar{\nu _{l}}$, where $P$
is one of the pseudoscalar mesons $\pi ^{+}$ and $K^{+}$. At the last stage $%
T_{bs;\overline{u}\overline{d}}^{-}$ should also decay through weak
processes and create a new tetraquark, which may be unstable or stable
against the strong interactions. Therefore, semileptonic decays of $T_{bs;%
\overline{u}\overline{d}}^{-}$ to ordinary mesons through intermediate
4-quark state are important for throughout analysis of the tetraquark $%
T_{bb}^{-}$.

In the present work we consider namely the processes $T_{bs;\overline{u}%
\overline{d}}^{-}\rightarrow T_{cs;\overline{u}\overline{d}}^{0}l\bar{\nu
_{l}}$, with $l=e,\mu ,$ and $\tau $ (in what follows we denote $T_{bs;%
\overline{u}\overline{d}}^{-}\Rightarrow T_{bs}^{-}$ and $T_{cs;\overline{u}%
\overline{d}}^{0}\Rightarrow T_{cs}^{0}$, respectively), and calculate their
partial widths. To this end, we first explore the properties of the scalar
4-quark state $T_{cs}^{0}$ and calculate its mass and coupling. Our
prediction for the mass of this state $m_{T}=\left( 2878\pm 128\right) ~%
\mathrm{MeV}$ demonstrates that $T_{cs}^{0}$ can decay strongly to the
conventional mesons $D^{0}\overline{K}^{0}$ and $D^{+}K^{-}$, partial widths
of which are computed as well. Using information on parameters of $%
T_{cs}^{0} $, we study the semileptonic decays of the tetraquark $T_{bs}^{-}$
and find branching ratios of the processes $T_{bs}^{-}\rightarrow D^{0}%
\overline{K}^{0}l\bar{\nu _{l}}$ and $T_{bs}^{-}\rightarrow D^{+}K^{-}l\bar{%
\nu _{l}}$. Results of the present work allow us also to analyze decays of
the tetraquark $T_{bb}^{-}$ and trace back its transformations to ordinary
mesons.

This paper is organized in the following manner. In Sec.\ \ref{sec:MassDecay}
we calculate the spectroscopic parameters of the scalar 4-quark state $%
T_{cs}^{0}$. Its strong decays are also analyzed in this section. The
section \ref{sec:Weak} is devoted to semileptonic decays, where we calculate
the weak form factors $G_{1(2)}(q^{2})$ and partial widths of the processes $%
T_{bs}^{-}\rightarrow T_{cs}^{0}l\bar{\nu _{l}}$. In Sec.\ \ref{sec:AC} we
sum up information on $T_{bs}^{-}$, and analyze transformations of $%
T_{bb}^{-}$ to conventional mesons.

%%%%%%%%%%%%%%%%%%%%%%%%%%%%%%%%%%%%%%%%%%%%%%%%%%%%%%%%%%%%%%%%%%%%%%%%%%%%%%%555

\section{Spectroscopic parameters and strong decays of the tetraquark $%
T_{cs}^{0}$}

\label{sec:MassDecay}
%%%%%%%%%%%%%%%%%%%%%%%%%%%%%%%%%%%%%%%%%%%%%%%%%%%%%%%%%%%

It has been emphasized above that transformation of the $T_{bs}^{-}$ to
meson pairs $D^{0}\overline{K}^{0}$ and $D^{+}K^{-}$ runs through creating
and decaying of the intermediate scalar 4-quark state $T_{cs}^{0}$. Hence,
parameters of this tetraquark are essential for our following analysis. In
this section we calculate the mass and coupling of the tetraquark $%
T_{cs}^{0} $ by means of the QCD two-point sum rule method, which is an
effective and powerful nonperturbative approach to investigate parameters of
hadrons \cite{Shifman:1978bx,Shifman:1978by}. It can be used to determine
masses, couplings, and decay widths not only of the conventional hadrons,
but also of exotic states\ \cite{Albuquerque:2018jkn}. In calculations, we
take into account effects of the vacuum condensates up to dimension 10.

Here, we also analyze decays of this exotic state to conventional mesons via
strong interactions. For these purposes, we use the parameters of the
tetraquark $T_{cs}^{0}$ and calculate the strong couplings $g_{TD^{0}%
\overline{K}^{0}}$ and $g_{TD^{+}K^{-}}$ corresponding to the vertices $%
T_{cs}^{0}D^{0}\overline{K}^{0}$and $T_{cs}^{0}D^{+}K^{-}$, respectively.
These couplings are necessary to find the widths of the $S$-wave decays $%
T_{cs}^{0}\rightarrow D^{0}\overline{K}^{0}$ and $T_{cs}^{0}\rightarrow
D^{+}K^{-}$, and can be calculated by means the QCD light-cone sum rule
(LCSR) approach \cite{Balitsky:1989ry}. Because the aforementioned vertices
contain a tetraquark the LCSR method should be supplemented by a technique
of the soft-meson approximation \cite{Belyaev:1994zk}. For investigation of
the diquark-antidiquark states the soft-meson approximation was adjusted in
Ref.\ \cite{Agaev:2016dev}, and successfully applied later to explore their
strong decays (see, for example, Refs.\ \cite%
{Agaev:2016ijz,Sundu:2017xct,Sundu:2018nxt}).

%%%%%%%%%%%%%%%%%%%%%%%%%%%%%%%%%%%%%%%%%%%%%%%%%%%%%%%%%%%%%%%%%%%%%%%%%%%

\subsection{Mass and coupling of the $T_{cs}^{0}$}

\label{subsec:MC}
%%%%%%%%%%%%%%%%%%%%%%%%%%%%%%%%%%%%%%%%%%%%%%%%%%%%%%%%%%%%%%%%%%%%%%%%%%%%%%
The mass and coupling of the tetraquark $T_{cs}^{0}$ can be obtained from
the QCD two-point sum rules. To this end, we start from the analysis of the
two-point correlation function
\begin{equation}
\Pi (p)=i\int d^{4}xe^{ipx}\langle 0|\mathcal{T}\{J^{T}(x)J^{T\dag
}(0)\}|0\rangle ,  \label{eq:CF}
\end{equation}%
where
\begin{equation}
J^{T}(x)=\epsilon \widetilde{\epsilon }[c_{b}^{T}(x)C\gamma _{5}s_{c}(x)][%
\overline{u}_{d}(x)\gamma _{5}C\overline{d}_{e}^{T}(x)]  \label{eq:Curr}
\end{equation}%
is the interpolating current for the tetraquark $T_{cs}^{0}$. Here, $%
\epsilon \widetilde{\epsilon }=\epsilon ^{abc}\epsilon ^{ade}$, and $a,b,c,d$%
, and $e$ are color indices and $C$ is the charge-conjugation operator.

We assume that $T_{cs}^{0}$ is composed of the scalar diquark $\epsilon
^{abc}[c_{b}^{T}C\gamma _{5}s_{c}]$ in the color antitriplet and flavor
antisymmetric state, and the antidiquark $\epsilon ^{ade}[\overline{u}%
_{d}\gamma _{5}C\overline{d}_{e}^{T}]$ in the color triplet state. Because
these diquark configurations are most attractive ones \cite{Jaffe:2004ph},
the current (\ref{eq:Curr}) corresponds to the ground-state scalar particle $%
T_{cs}^{0}$ with lowest mass.

To find the phenomenological side of the sum rule $\Pi ^{\mathrm{Phys}}(p)$,
we use the "ground-state+continuum" scheme. Then, $\Pi ^{\mathrm{Phys}}(p)$
contains a contribution of the ground-state particle which below is written
down explicitly, and effects of higher resonances and continuum states
denoted by dots
\begin{equation}
\Pi ^{\mathrm{Phys}}(p)=\frac{\langle 0|J|T_{cs}^{0}(p)\rangle \langle
T_{cs}^{0}(p)|J^{\dagger }|0\rangle }{m_{T}^{2}-p^{2}}+\ldots
\label{eq:Phys1}
\end{equation}%
\qquad

The QCD side of the sum rules is determined by the same correlation function
$\Pi ^{\mathrm{OPE}}(p)$ found using the perturbative QCD and expressed in
terms of the quark propagators. Expressions for the invariant amplitudes $%
\Pi ^{\mathrm{Phys}}(p^{2})$ and $\Pi ^{\mathrm{OPE}}(p^{2})$, which are
necessary to derive the required sum rules for the mass $m_{T}$ and coupling
$f_{T}$ of the tetraquark $T_{cs}^{0}$, as well as manipulations with these
functions are similar to ones presented in Ref.\ \cite{Sundu:2019feu},
therefore we do not repeat them here; required theoretical results can be
obtained from corresponding expressions for the $T_{bs}^{-}$ by a simple $%
b\rightarrow c$ replacement.

The sum rules for $m_{T}$ and $f_{T}$ contain the quark, gluon and mixed
vacuum condensates, values of which are collected in Table \ref{tab:PM}.
This table contains also the masses of the $b,\ c,$ and $s$ quarks, as well
as spectroscopic parameters of the mesons $D$ and $K$, which will be
utilized in the next subsection.
\begin{table}[tbp]
\begin{tabular}{|c|c|}
\hline\hline
Quantity & Value \\ \hline\hline
$\langle \bar{q}q \rangle $ & $-(0.24\pm 0.01)^3~\mathrm{GeV}^3$ \\
$\langle \bar{s}s \rangle $ & $0.8\langle \bar{q}q \rangle$ \\
$m_{0}^2 $ & $(0.8\pm0.1)~\mathrm{GeV}^2$ \\
$\langle \overline{s}g_{s}\sigma Gs\rangle$ & $m_{0}^2\langle \bar{s}s
\rangle $ \\
$\langle\frac{\alpha_sG^2}{\pi}\rangle $ & $(0.012\pm0.004)~\mathrm{GeV}^4$
\\
$\langle g_{s}^3G^3\rangle $ & $(0.57\pm0.29)~\mathrm{GeV}^6 $ \\
$m_{b}$ & $(4.18 \pm 0.03)~\mathrm{GeV}$ \\
$m_{c}$ & $(1.275 \pm 0.025)~\mathrm{GeV}$ \\
$m_{s} $ & $93^{+11}_{-5}~\mathrm{MeV} $ \\
$m_{K^{0}}$ & $(497.614 \pm 0.024)~\mathrm{MeV}$ \\
$m_{K^{-}}$ & $(493.677 \pm 0.016)~\mathrm{MeV}$ \\
$m_{D}$ & $(1864.84 \pm 0.07)~\mathrm{MeV}$ \\
$m_{D^{+}}$ & $(1869.61 \pm 0.10)~\mathrm{MeV}$ \\
$f_{K^{-}}=f_{K^{0}}$ & $(155.72 \pm 0.51)~\mathrm{MeV}$ \\
$f_{D}=f_{D^{+}}$ & $(203.7 \pm 4.7)~\mathrm{MeV}$ \\ \hline\hline
\end{tabular}%
\caption{Parameters used in calculations.}
\label{tab:PM}
\end{table}

The sum rules also depend on two auxiliary parameters. First of them is $%
M^{2}$, which appears in expressions after applying the Borel transformation
to sum rules to suppress contributions of the higher resonances and
continuum states. The dependence on the continuum threshold parameter $s_{0}$
is an output of the continuum subtraction procedure. A choice of these
parameters is controlled by constraints on the pole contribution ($\mathrm{PC%
}$) and convergence of the operator product expansion ($\mathrm{OPE}$), as
well as by a minimum sensitivity of the extracted quantities on $M^{2}$ and $%
s_{0}$.

Thus, the maximum allowed $M^{2}$ should be fixed to obey the restriction
imposed on $\mathrm{PC}$
\begin{equation}
\mathrm{PC}=\frac{\Pi (M^{2},\ s_{0})}{\Pi (M^{2},\ \infty )},  \label{eq:PC}
\end{equation}%
where $\Pi (M^{2},\ s_{0})$ is the Borel-transformed and subtracted
invariant amplitude $\Pi ^{\mathrm{OPE}}(p^{2})$. The lower bound of the
window for the Borel parameter is determined from convergence of the $%
\mathrm{OPE}$, which can be quantified by the ratio%
\begin{equation}
R(M^{2})=\frac{\Pi ^{\mathrm{DimN}}(M^{2},\ s_{0})}{\Pi (M^{2},\ s_{0})}.
\label{eq:Convergence}
\end{equation}%
Here $\Pi ^{\mathrm{DimN}}(M^{2},\ s_{0})$ denotes a contribution to the
correlation function of the last term (or a sum of last few terms) in the
operator product expansion. A stability of extracted quantities is among
important requirements of the sum rule calculations.

In the present work, at the maximum of $M^{2}$ we apply the constraint $%
\mathrm{PC}>0.2$ which is typical for multiquark systems. To ensure
convergence of the $\mathrm{OPE}$, at the minimum limit of $M^{2}$ we use
the restriction $R\leq 0.01$. Performed analysis demonstrates that the
working regions
\begin{equation}
M^{2}\in \lbrack 1.8,\ 2.8]~\mathrm{GeV}^{2},\ s_{0}\in \lbrack 11,\ 12]~%
\mathrm{GeV}^{2},  \label{eq:Wind1}
\end{equation}%
obey the constraints imposed on the Borel and continuum threshold
parameters. Indeed, the pole contribution at $M^{2}=\ 2.8\ \mathrm{GeV}^{2}$
amounts to $\mathrm{PC}=0.22$, whereas at $M^{2}=1.8~\mathrm{GeV}^{2}$ it
reaches the maximum value $0.61$. Numerical computations show that for $%
\mathrm{DimN}=\mathrm{Dim(8+9+10)}$ the ratio $R(1.8~\mathrm{GeV}^{2})$ is
equal to $0.007$, which guarantees the convergence of the sum rules. These
two values of $M^{2}$ determine the boundaries of the region within of which
the Borel parameter can be varied.

In general, quantities extracted from sum rules should not depend on the
auxiliary parameters used in calculations. In real computations, however,
these quantities, i.e., $m_{T}$ and $f_{T}$ in the case under consideration,
demonstrate a residual dependence on $M^{2}$ and $s_{0}$. Let us note that a
dependence on the parameters $M^{2}$ and $s_{0}$ is a main source of
unavoidable theoretical errors in the sum rule calculations, which however
can be systematically taken into account.
\begin{figure}[h]
\includegraphics[width=8.8cm]{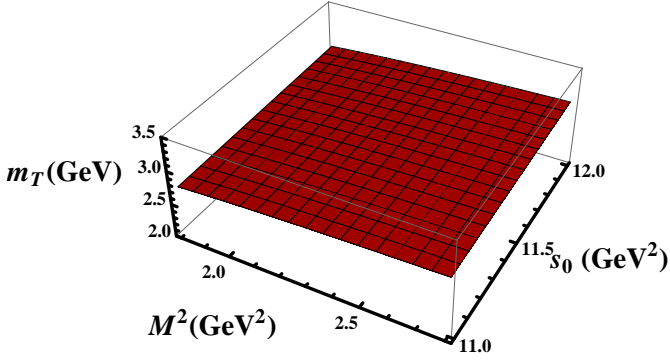}
\caption{The mass $m_{T}$ of the tetraquark $T_{cs}^{0}$ as a function of
the Borel and continuum threshold parameters.}
\label{fig:Mass}
\end{figure}

\begin{figure}[h]
\includegraphics[width=8.8cm]{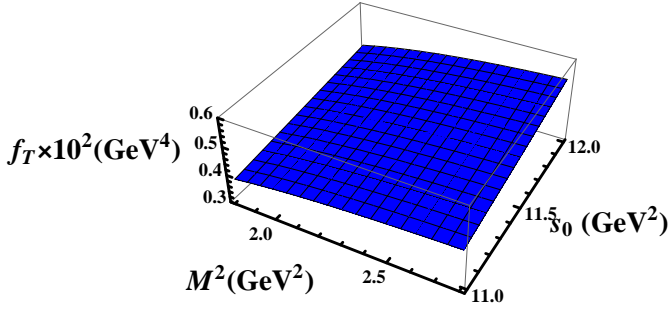}
\caption{The same as in Fig.\ 1, but for the coupling $f_{T}$.}
\label{fig:Coupl}
\end{figure}
In Figs.\ \ref{fig:Mass} and \ref{fig:Coupl} we plot the predictions for the
mass $m_{T}$ and coupling $f_{T}$, in which one can see their dependence on
the parameters $M^{2}$ and $s_{0}$.

Our results for the spectroscopic parameters of the tetraquark $T_{cs}^{0}$
read
\begin{eqnarray}
m_{T} &=&\left( 2878\pm 128\right) ~\mathrm{MeV},\   \notag \\
f_{T} &=&\left( 0.45\pm 0.08\right) \times 10^{-2}~\mathrm{GeV}^{4}.
\label{eq:Result1}
\end{eqnarray}%
These predictions will be used below to study the strong decays of $%
T_{cs}^{0}$.

%%%%%%%%%%%%%%%%%%%%%%%%%%%%%%%%%%%%%%%%%%%%%%%%%%%%%%%%%%%%%%%%%%%%%%%%%%%%

\subsection{Strong decays $T_{cs}^{0}\rightarrow D^{0}\overline{K}^{0}$ and $%
T_{cs}^{0}\rightarrow D^{+}K^{-}$}

\label{subsec:Decays}
%%%%%%%%%%%%%%%%%%%%%%%%%%%%%%%%%%%%%%%%%%%%%%%%%%%%%%%%%%%%%%%%%%%%%%%%
The spectroscopic parameters of the tetraquark $T_{cs}^{0}$ obtained in the
previous subsection provide an information necessary to answer a question
about its stability against the strong interactions. It is not difficult to
see, that the mass $m_{T}$ makes kinematically allowed the strong decays $%
T_{cs}^{0}\rightarrow D^{0}\overline{K}^{0}$ and $T_{cs}^{0}\rightarrow
D^{+}K^{-}$. There are other strong decay modes of $T_{cs}^{0}$, but these
two channels are $S$-wave processes. Here, we are going to consider in a
detailed form the channel $T_{cs}^{0}\rightarrow D^{0}\overline{K}^{0}$, and
give final results for the second one.

The width of the decay $T_{cs}^{0}\rightarrow D^{0}\overline{K}^{0}$, apart
from other parameters, is determined by the strong coupling $g_{TD^{0}%
\overline{K}^{0}}$ corresponding to the vertex $T_{cs}^{0}D^{0}\overline{K}%
^{0}$. Our aim is to calculate $g_{TD^{0}\overline{K}^{0}}$ which
quantitatively describes strong interactions between the tetraquark and two
conventional mesons. To this end, we use the LCSR method and begin from
analysis of the correlation function
\begin{equation}
\Pi (p,q)=i\int d^{4}xe^{ipx}\langle \overline{K}^{0}(q)|\mathcal{T}%
\{J^{D^{0}}(x)J^{T\dagger }(0)\}|0\rangle ,  \label{eq:CFA}
\end{equation}%
where $J^{D^{0}}(x)$ is the interpolating current of the meson $D^{0}$; it
has following form
\begin{equation}
J^{D^{0}}(x)(x)=\overline{u}(x)i\gamma _{5}c(x).  \label{eq:CurrA}
\end{equation}%
Standard recipes require to write $\Pi (p,q)$ in terms of physical
parameters of the particles $T_{cs}^{0}$, $D^{0}$, and $\overline{K}^{0}$
\begin{eqnarray}
&&\Pi ^{\mathrm{Phys}}(p,q)=\frac{\langle 0|J^{D^{0}}|D^{0}(p)\rangle }{%
p^{2}-m_{D}^{2}}\langle D^{0}\left( p\right) \overline{K}%
^{0}(q)|T_{cs}^{0}(p^{\prime })\rangle  \notag \\
&&\times \frac{\langle T_{cs}^{0}(p^{\prime })|J^{T\dagger }|0\rangle }{%
p^{\prime 2}-m_{T}^{2}}+...,  \label{eq:PhysDec2}
\end{eqnarray}%
where $p^{\prime }$ and $p$, $q$ are 4-momenta of the initial and final
particles, respectively. In the expression above by dots we note
contributions of excited resonances and continuum states. The correlation
function $\Pi ^{\mathrm{Phys}}(p,q)$ can be simplified by introducing the
matrix elements
\begin{eqnarray}
&&\langle 0|J^{D^{0}}|D^{0}(p)\rangle =\frac{f_{D}m_{D}^{2}}{m_{c}+m_{u}},
\notag \\
&&\langle D^{0}\left( p\right) \overline{K}^{0}(q)|T_{cs}^{0}(p^{\prime
})\rangle =g_{TD^{0}\overline{K}^{0}}(p\cdot p^{\prime }).  \label{eq:ME1}
\end{eqnarray}%
The matrix element $\langle 0|J^{D^{0}}|D^{0}(p)\rangle $ is expressed in
terms of $D^{0}$ meson's mass $m_{D}$ and its decay constant $f_{D}$,
whereas $\langle D^{0}\left( p\right) \overline{K}^{0}(q)|T_{cs}^{0}(p^{%
\prime })\rangle $ is written down using the strong coupling $g_{TD^{0}%
\overline{K}^{0}}$. In the soft-meson limit $q\rightarrow 0$ we get $%
p^{\prime }=p$ \cite{Agaev:2016dev}, and must carry out the Borel
transformation of $\Pi ^{\mathrm{Phys}}(p,q=0)$ over the variable $p^{2}$,
which gives
\begin{equation}
\mathcal{B}\Pi ^{\mathrm{Phys}}(p^{2})=g_{TD^{0}\overline{K}^{0}}\frac{%
f_{D}m_{D}^{2}f_{T}m_{T}\widetilde{m}^{2}}{m_{c}+m_{u}}\frac{e^{-\widetilde{m%
}^{2}/M^{2}}}{M^{2}}\ldots ,  \label{eq:BorelPhys}
\end{equation}%
where
\begin{equation}
\widetilde{m}^{2}=\frac{m_{T}^{2}+m_{D}^{2}}{2}.
\end{equation}

The necessity to use the soft-meson approximation of the LCSR method and set
$q=0$ is connected with features of tetraquark-meson-meson strong vertices.
Because a tetraquark is built of four valence quarks, calculations of the
correlation function (\ref{eq:CFA}) by contracting quark fields from
relevant interpolating currents lead to appearance of two quark fields at
the same space-time position, which, sandwiched between the vacuum and $%
\overline{K}^{0}$ meson, generate the local matrix elements of $\overline{K}%
^{0}$. Then, to preserve the 4-momentum conservation at the vertex one has
to set $q=0$, and employ technical tools of soft-meson approach elaborated
in the full LCSR method as the approximation to vertices containing only
conventional mesons \cite{Belyaev:1994zk}. Let us emphasize that in the case
of tetraquark-meson-meson vertices soft limit is an only way to calculate
corresponding strong couplings in the framework of the LCSR method.

The soft approximation modifies the physical side of the sum rules. A
problem is that in the soft limit some of contributions arising from the
higher resonances and continuum states even after the Borel transformation
remain unsuppressed. These terms correspond to vertices containing excited
states of involved particles, and contaminate the physical side of sum
rules. Therefore, before performing the continuum subtraction in the final
sum rule they should be delated by means of some manipulations. This problem
can be solved by acting on the physical side of the sum rule by the operator
\cite{Belyaev:1994zk,Ioffe:1983ju}
\begin{equation}
\mathcal{P}(M^{2},\widetilde{m}^{2})=\left( 1-M^{2}\frac{d}{dM^{2}}\right)
M^{2}e^{\widetilde{m}^{2}/M^{2}},  \label{eq:Oper}
\end{equation}%
which keeps unchanged the ground-state term removing, at the same time,
unsuppressed contributions. Naturally, the operator $\mathcal{P}(M^{2},%
\widetilde{m}^{2})$ has to be applied to the QCD side of the sum rule as
well, which has to be calculated in the soft-meson approximation and
expressed in terms of the $\overline{K^{0}}$ meson's local matrix elements.

In the soft limit the correlation function $\Pi ^{\mathrm{OPE}}(p)$ is
determined by the expression
\begin{eqnarray}
&&\Pi ^{\mathrm{OPE}}(p)=i\int d^{4}xe^{ipx}\epsilon \widetilde{\epsilon }%
\left[ \gamma _{5}\widetilde{S}_{c}^{ib}(x){}\gamma
_{5}S_{u}^{di}(-x){}\gamma _{5}\right] _{\alpha \beta }  \notag \\
&&\times \langle \overline{K}^{0}|\overline{s}_{\alpha }^{c}(0)d_{\beta
}^{e}(0)|0\rangle ,  \label{eq:OPE1}
\end{eqnarray}%
where%
\begin{equation}
\widetilde{S}(x)=CS_{c(q)}^{T}(x)C.  \label{eq:Prop}
\end{equation}%
In Eqs.\ (\ref{eq:OPE1}) and (\ref{eq:Prop}), $S_{c(q)}(x)$ are the $c$
quark and light quark propagators explicit expressions of which can be found
in Ref.\ \cite{Sundu:2018uyi}; for simplicity we do not provide these
formulas here.

As is seen, the correlation function \ $\Pi ^{\mathrm{OPE}}(p)$ depends on
local matrix elements $\langle \overline{K}^{0}|\overline{s}_{\alpha
}^{c}(0)d_{\beta }^{e}(0)|0\rangle $, which should be recast to forms
suitable for expressing them as standard matrix elements of $\overline{K}%
^{0} $. For these purposes, we employ the expansion
\begin{equation}
\overline{s}_{\alpha }^{c}d_{\beta }^{e}\rightarrow \frac{1}{12}\Gamma
_{\beta \alpha }^{j}\delta ^{ce}\left( \overline{s}\Gamma ^{j}d\right) ,
\label{eq:MatEx}
\end{equation}%
where $\Gamma ^{j}$ is the full set of Dirac matrices
\begin{equation}
\Gamma ^{j}=\mathbf{1,\ }\gamma _{5},\ \gamma _{\lambda },\ i\gamma
_{5}\gamma _{\lambda },\ \sigma _{\lambda \rho }/\sqrt{2}.
\end{equation}%
Then operators $\overline{s}\Gamma ^{j}d$ and ones appeared due to $G$
insertions from propagators $\widetilde{S}$ and $S$, give rise to local
matrix elements of the $\overline{K}^{0}$ meson. Substituting Eq.\ (\ref%
{eq:MatEx}) into the correlation function and performing the color summation
in accordance with prescriptions described in Ref.\ \cite{Agaev:2016dev}, we
fix twist-3 local matrix element of $\overline{K}^{0}$
\begin{equation}
\langle 0|\overline{d}(0)i\gamma _{5}s(0)|\overline{K}^{0}\rangle =\frac{%
f_{K^{0}}m_{K^{0}}^{2}}{m_{s}+m_{d}},  \label{eq;ME1}
\end{equation}%
that contributes to the correlation function.

The function $\Pi ^{\mathrm{OPE}}(p)$ contains the trivial Lorentz structure
which is proportional to $I$. The Borel transformed and subtracted
expression of the corresponding invariant amplitude $\Pi ^{\mathrm{OPE}%
}(p^{2})$ reads%
\begin{eqnarray}
&&\Pi ^{\mathrm{OPE}}(M^{2},s_{0})=\int_{(m_{c}+m_{s})^{2}}^{s_{0}}ds\rho ^{%
\mathrm{pert.}}(s)e^{-s/M^{2}}  \notag \\
&&+\frac{\mu _{K^{0}}}{6}e^{-m_{c}^{2}/M^{2}}\left\{ m_{c}\langle \overline{q%
}q\rangle +\frac{1}{8}\langle \frac{\alpha _{s}G^{2}}{\pi }\rangle \left[ 1+%
\frac{m_{c}^{2}}{6M^{2}}\right] \right.  \notag \\
&&-\frac{m_{c}^{3}}{4M^{4}}\langle \overline{s}g_{s}\sigma Gs\rangle -\frac{%
g_{s}^{2}m_{c}^{4}}{81M^{6}}\langle \overline{q}q\rangle ^{2}  \notag \\
&&\left. -\frac{m_{c}\pi ^{2}}{18M^{6}}\langle \frac{\alpha _{s}G^{2}}{\pi }%
\rangle \langle \overline{q}q\rangle \left( m_{c}^{2}-3M^{2}\right) \right\}
,  \label{eq:Borel}
\end{eqnarray}%
where
\begin{equation}
\rho ^{\mathrm{pert.}}(s)=\frac{\mu _{K^{0}}}{24\pi ^{2}}(3m_{c}^{2}-s),
\label{eq:Pert}
\end{equation}%
and $\mu _{K^{0}}=f_{K^{0}}m_{K^{0}}^{2}/(m_{s}+m_{d})$. Let us note that
calculations of $\Pi ^{\mathrm{OPE}}(M^{2},s_{0})$ are carried out by taking
into account nonperturbative terms up to seventh dimension. Then, the sum
rule for the strong coupling $g_{TD^{0}\overline{K}^{0}}$ takes the form%
\begin{equation}
g_{TD^{0}\overline{K}^{0}}=\frac{m_{c}+m_{u}}{f_{D}m_{D}^{2}f_{T}m_{T}%
\widetilde{m}^{2}}\mathcal{P}(M^{2},\widetilde{m}^{2})\Pi ^{\mathrm{OPE}%
}(M^{2},s_{0}).  \label{eq:SR}
\end{equation}%
The width of the decay $T_{cs}^{0}\rightarrow D^{0}\overline{K}^{0}$ is
given by the formula%
\begin{equation}
\Gamma \lbrack T_{cs}^{0}\rightarrow D^{0}\overline{K}^{0}]=\frac{g_{TD^{0}%
\overline{K}^{0}}^{2}m_{D}^{2}}{8\pi }\lambda \left( 1+\frac{\lambda ^{2}}{%
m_{D}^{2}}\right) ,  \label{eq:DW1}
\end{equation}%
where%
\begin{eqnarray}
\lambda &=&\lambda \left( m_{T}^{2},m_{D}^{2},m_{K^{0}}^{2}\right) =\frac{1}{%
2m_{T}}\left[ m_{T}^{4}+m_{K^{0}}^{4}+m_{D}^{4}\right.  \notag \\
&&\left.
-2(m_{T}^{2}m_{D}^{2}+m_{T}^{2}m_{K^{0}}^{2}+m_{K^{0}}^{2}m_{D}^{2}) \right]
^{1/2}.  \label{eq:Lambda}
\end{eqnarray}

\begin{figure}[h]
\includegraphics[width=8.8cm]{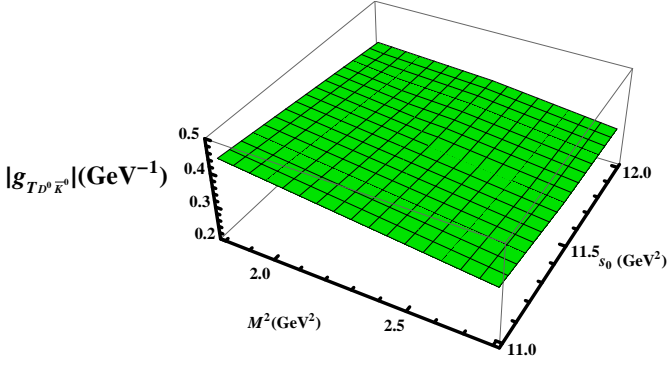}
\caption{The strong coupling $|g_{TD^{0}\overline{K}^{0}}|$ as a function of
the Borel and continuum threshold parameters.}
\label{fig:StCoupl}
\end{figure}

In numerical computations of $g_{TD^{0}\overline{K}^{0}}$ the Borel and
continuum threshold parameters are chosen as in Eq.\ (\ref{eq:Wind1}). To
visualize a sensitivity of the strong coupling on these parameters, in Fig.\ %
\ref{fig:StCoupl} we depict the dependence of $|g_{TD^{0}\overline{K}^{0}}|$
on $M^{2}$ and $s_{0}$; ambiguities generated by the choice of these
parameters do not exceed \ $\pm 19\%$ of the central value.

For the strong coupling $g_{TD^{0}\overline{K}^{0}}$ our analysis yields%
\begin{equation}
|g_{TD^{0}\overline{K}^{0}}|=(0.37\pm 0.07)~\mathrm{GeV}^{-1}.
\label{eq:SC1}
\end{equation}%
Using the result obtained for $g_{TD^{0}\overline{K}^{0}}$, we can evaluate
the partial width of the decay $T_{cs}^{0}\rightarrow D^{0}\overline{K}^{0}$:%
\begin{equation}
\Gamma \left( T_{cs}^{0}\rightarrow D^{0}\overline{K}^{0}\right) =(15.35\pm
4.11)~\mathrm{MeV.}  \label{eq:NDW1}
\end{equation}

The decay $T_{cs}^{0}\rightarrow D^{+}K^{-}$ can be analyzed by the same
manner. The difference is connected with quark contents of the mesons $D^{+}$
and $K^{-}$ that generate small modifications, for example, $\widetilde{\Pi }%
^{\mathrm{OPE}}(p)$ takes the form
\begin{eqnarray}
\widetilde{\Pi }^{\mathrm{OPE}}(p) &=&i\int d^{4}xe^{ipx}\epsilon \widetilde{%
\epsilon }\left[ \gamma _{5}\widetilde{S}_{c}^{ib}(x){}\gamma
_{5}S_{d}^{di}(-x){}\gamma _{5}\right] _{\alpha \beta }  \notag \\
&&\times \langle K^{-}|\overline{u}_{\alpha }^{c}(0)s_{\beta
}^{e}(0)|0\rangle .  \label{eq:OPE2}
\end{eqnarray}%
Therefore, we write down the final results for the strong coupling $%
g_{TD^{+}K^{-}}$ and corresponding decay width
\begin{eqnarray}
&&|g_{TD^{+}K^{-}}|=(0.38\pm 0.06)\ \mathrm{GeV}^{-1},  \notag \\
&&\Gamma \left( T_{cs}^{0}\rightarrow D^{+}K^{-}\right) =(15.40\pm 3.44)~%
\mathrm{MeV}.  \label{eq:DW2}
\end{eqnarray}%
These dominant decay channels allow us to estimate the full width of the
tetraquark $T_{cs}^{0}$%
\begin{equation}
\Gamma =(30.8\pm 5.4)~\mathrm{MeV.}
\end{equation}%
In light of obtained prediction for the full width of $T_{cs}^{0}$, we
classify it as a relatively narrow unstable tetraquark.

%%%%%%%%%%%%%%%%%%%%%%%%%%%%%%%%%%%%%%%%%%%%%%%%%%%%%%%%%%%

\section{Semileptonic decay $T_{bs}^{-}\rightarrow T_{cs}^{0}l\overline{%
\protect\nu }_{l}$}

\label{sec:Weak}
%%%%%%%%%%%%%%%%%%%%%%%%%%%%%%%%%%%%%%%%%%%%%%%%%%%%%%%%%%%%%%%

The semileptonic decay $T_{bs}^{-}\rightarrow T_{cs}^{0}l\overline{\nu }_{l}$
runs through the transitions $b\rightarrow W^{-}c$ and $W^{-}\rightarrow l%
\overline{\nu _{l}}$. It is not difficult to see that decays with all lepton
species $l=e,\ \mu $ and $\tau $ are kinematically allowed processes.

The transition $b\rightarrow c$ at the tree-level can be described using the
effective Hamiltonian
\begin{equation}
\mathcal{H}^{\mathrm{eff}}=\frac{G_{F}}{\sqrt{2}}V_{bc}\overline{c}\gamma
_{\mu }(1-\gamma _{5})b\overline{l}\gamma ^{\mu }(1-\gamma _{5})\nu _{l},
\label{eq:EffH}
\end{equation}%
where $G_{F}$ is the Fermi coupling constant, and $V_{bc}$ is the relevant
Cabibbo-Kobayashi-Maskawa (CKM) matrix element. After placing the effective
Hamiltonian $\mathcal{H}^{\mathrm{eff}}$ between the initial and final
tetraquarks and factoring out the lepton fields one gets the matrix element
of the current%
\begin{equation}
J_{\mu }^{\mathrm{W}}=\overline{c}\gamma _{\mu }(1-\gamma _{5})b.
\label{eq:TrCurr}
\end{equation}%
The matrix element $\langle T_{cs}^{0}(p^{\prime })|J_{\mu }^{\mathrm{W}%
}|T_{bs}^{-}(p)\rangle $ can be expressed in terms of the form factors $%
G_{i}(q^{2})$ that parameterize the long-distance dynamics of the weak
transition. In the case of scalar tetraquarks it has the rather simple form
\begin{equation}
\langle T_{cs}^{0}(p^{\prime })|J_{\mu }^{\mathrm{W}}|T_{bs}^{-}(p)\rangle
=G_{1}(q^{2})P_{\mu }+G_{2}(q^{2})q_{\mu },  \label{eq:Vertex1}
\end{equation}%
where $p$ and $p^{\prime }$ are the momenta of the tetraquarks $T_{bs}^{-}$
and $T_{cs}^{0}$, respectively. Here, we use the shorthand notations $P_{\mu
}=p_{\mu }^{\prime }+p_{\mu }$ and $q_{\mu }=p_{\mu }-p_{\mu }^{\prime }$.
The $q_{\mu }$ is the momentum transferred to the leptons, and $q^{2}$
changes within the limits $m_{l}^{2}\leq q^{2}\leq (m-m_{T})^{2}$, where $%
m_{l}$ is the mass of a lepton $l$.

The sum rules for the form factors $G_{i}(q^{2}),i=1,2$ can be derived from
the three-point correlation function
\begin{eqnarray}
\Pi _{\mu }(p,p^{\prime }) &=&i^{2}\int d^{4}xd^{4}ye^{i(p^{\prime }y-px)}
\notag \\
&&\times \langle 0|\mathcal{T}\{J^{T}(y)J_{\mu }^{\mathrm{W}}(0)J^{\dagger
}(x)\}|0\rangle ,  \label{eq:CF2}
\end{eqnarray}%
where $J^{T}(y)$ and $J(x)$ are the interpolating currents for the states $%
T_{cs}^{0}$ and $T_{bs}^{-}$, respectively. The current $J^{T}(y)$ is given
by Eq.\ (\ref{eq:Curr}), whereas for $J(x)$ we use the expression
\begin{equation}
J(x)=\epsilon \widetilde{\epsilon }[b_{b}^{T}(x)C\gamma _{5}s_{c}(x)][%
\overline{u}_{d}(x)\gamma _{5}C\overline{d}_{e}^{T}(x)].
\end{equation}

First, we express the correlation function $\Pi _{\mu }(p,p^{\prime })$ in
terms of the spectroscopic parameters of the tetraquark and mesons, and fix
the physical side of the sum rule, i.e., find the function $\Pi _{\mu }^{%
\mathrm{Phys}}(p,p^{\prime })$. It can be easily written down in the form%
\begin{eqnarray}
&&\Pi _{\mu }^{\mathrm{Phys}}(p,p^{\prime })=\frac{\langle
0|J^{T}|T_{cs}^{0}(p^{\prime })\rangle \langle T_{cs}^{0}(p^{\prime
})|J_{\mu }^{\mathrm{W}}|T_{bs}^{-}(p)\rangle }{(p^{2}-m^{2})(p^{\prime
2}-m_{T}^{2})}  \notag \\
&&\times \langle T_{bs}^{-}(p)|J^{\dagger }|0\rangle +\ldots ,
\label{eq:Phys2}
\end{eqnarray}%
where we take explicitly into account a contribution of the ground-state
particles, and denote by dots effects due to excited and continuum states.

Using the tetraquarks' matrix elements and expressing the vertex $\langle
T(p^{\prime })|J_{\mu }^{\mathrm{W}}|T_{bs}^{-}(p)\rangle $ in terms of the
weak transition form factors $G_{i}(q^{2})$ it is not difficult to find that%
\begin{eqnarray}
\Pi _{\mu }^{\mathrm{Phys}}(p,p^{\prime }) &=&\frac{f_{T}m_{T}fm}{%
(p^{2}-m^{2})(p^{\prime 2}-m_{T}^{2})}  \notag \\
&&\times \left[ G_{1}(q^{2})P_{\mu }+G_{2}(q^{2})q_{\mu }\right] ,
\label{eq:Phys3}
\end{eqnarray}%
where the matrix element of the state $T_{bs}^{-}$ is defined by
\begin{equation}
\langle T_{bs}^{-}(p)|J^{\dagger }|0\rangle =fm.  \label{eq:ME3}
\end{equation}

To calculate $\Pi _{\mu }(p,p^{\prime })$, we employ the interpolating
currents and quark propagators, and find
\begin{eqnarray}
&&\Pi _{\mu }^{\mathrm{OPE}}(p,p^{\prime })=i^{2}\int
d^{4}xd^{4}ye^{i(p^{\prime }y-px)}\epsilon \widetilde{\epsilon }\epsilon
^{\prime }\widetilde{\epsilon }^{\prime }  \notag \\
&&\times \mathrm{Tr}\left[ \gamma _{5}\widetilde{S}_{d}^{e^{\prime
}e}(x-y)\gamma _{5}S_{u}^{d^{\prime }d}(x-y)\right] \mathrm{Tr}\left[ \gamma
_{\mu }(1-\gamma _{5})\right.  \notag \\
&&\left. \times S_{b}^{ib}(-x)\gamma _{5}\widetilde{S}_{s}^{cc^{\prime
}}(y-x)\gamma _{5}S_{c}^{b^{\prime }i}(y)\right] .  \label{eq:CF4}
\end{eqnarray}%
Then, the sum rules for the form factors $G_{i}(q^{2})$ can be derived by
equating the invariant amplitudes corresponding to the same Lorentz
structures in $\Pi _{\mu }^{\mathrm{Phys}}(p,p^{\prime })$ and $\Pi _{\mu }^{%
\mathrm{OPE}}(p,p^{\prime })$. Afterwards, we carry out the double Borel
transformation over $p^{\prime 2}$ and $p^{2}$ which is required to suppress
contributions of the higher excited and continuum states, and perform the
continuum subtraction. These operations lead to the sum rules
\begin{eqnarray}
&&G_{i}(\mathbf{M}^{2},\mathbf{s}_{0},q^{2})=\frac{1}{f_{T}m_{T}fm}%
\int_{(m_{b}+m_{s})^{2}}^{s_{0}}ds  \notag \\
&&\times \int_{(m_{c}+m_{s})^{2}}^{s_{0}^{\prime }}ds^{\prime }\rho
_{i}(s,s^{\prime },q^{2})e^{(m^{2}-s)/M_{1}^{2}}e^{(m_{T}^{2}-s^{\prime
})/M_{2}^{2}},  \notag \\
&&  \label{eq:FF}
\end{eqnarray}%
where $\rho _{1(2)}(s,s^{\prime },q^{2})$ are the spectral densities
calculated as the imaginary parts of the correlation function $\Pi _{\mu }^{%
\mathrm{OPE}}(p,p^{\prime })$ with dimension-five accuracy. In Eq.\ (\ref%
{eq:FF}) $\mathbf{M}^{2}$ and $\mathbf{s}_{0}$ are a couple of the Borel and
continuum threshold parameters, respectively; the set $(M_{1}^{2},s_{0})$
corresponds to the initial state $T_{bs}^{-}$, and the second set $%
(M_{2}^{2},s_{0}^{\prime })$ describes the tetraquark $T_{cs}^{0}$.

Parameters for numerical computations of $G_{i}(\mathbf{M}^{2},\mathbf{s}%
_{0},q^{2})$ are listed in Table \ref{tab:PM}. The mass and coupling of the
tetraquark $T_{bs}^{-}$
\begin{eqnarray}
m &=&(5380~\pm 170)~\mathrm{MeV},  \notag \\
f &=&(2.1\pm 0.5)\times 10^{-3}~\mathrm{GeV}^{4},  \label{eq:Initial}
\end{eqnarray}%
and working windows for the parameters $(M_{1}^{2},s_{0})$
\begin{equation}
M_{1}^{2}\in \lbrack 3.4,\ 4.8]~\mathrm{GeV}^{2},\ s_{0}\in \lbrack 35,\ 37]~%
\mathrm{GeV}^{2}  \label{eq:Wind2}
\end{equation}%
are borrowed from Ref. \cite{Sundu:2019feu}. The regions for $%
(M_{2}^{2},s_{0}^{\prime })$ and spectroscopic parameters of $T_{cs}^{0}$
are given by Eqs.\ (\ref{eq:Wind1}) and (\ref{eq:Result1}), respectively. In
numerical computations we also use the Fermi coupling constant $%
G_{F}=1.16637\times 10^{-5}~\mathrm{GeV}^{-2}$ and CKM matrix element $%
|V_{bc}|=(41.2\pm 1.01)\times 10^{-3}$. Like all quantities extracted from
sum rule computations, the weak form factors $G_{1(2)}(q^{2})$ depend on the
Borel and continuum threshold parameters $\mathbf{M}^{2}$ and $\mathbf{s}%
_{0}.$ Ambiguities connected with the choice of $(\mathbf{M}^{2},\mathbf{s}%
_{0})$ and ones due to other input parameters form theoretical errors of the
sum rule analysis, which will be taken into account in the fit functions.

To obtain the width of the decay $T_{bs}^{-}\rightarrow T_{cs}^{0}l\overline{%
\nu }_{l}$\ one must integrate the differential decay rate $d\Gamma /dq^{2}$
(see, explanation below) of this process in the kinematical limits $%
m_{l}^{2}\leq q^{2}\leq (m-m_{T})^{2}$. In the interval $m_{l}^{2}\leq
q^{2}\leq 5$ $\mathrm{GeV}^{2}$ the QCD sum rules lead to reliable
predictions for the form factors $G_{i}(q^{2})$ , which do not cover the
whole integration region $m_{l}^{2}\leq q^{2}\leq 6.26$ $\mathrm{GeV}^{2}$.
Therefore, we replace the weak form factors $G_{i}(q^{2})$ by the fit
functions $\mathcal{G}_{i}(q^{2})$, which at $q^{2}$ accessible for the sum
rule computations coincide with $G_{i}(q^{2})$, but can be easily
extrapolated to the full integration region.
\begin{figure}[h]
\includegraphics[width=8.8cm]{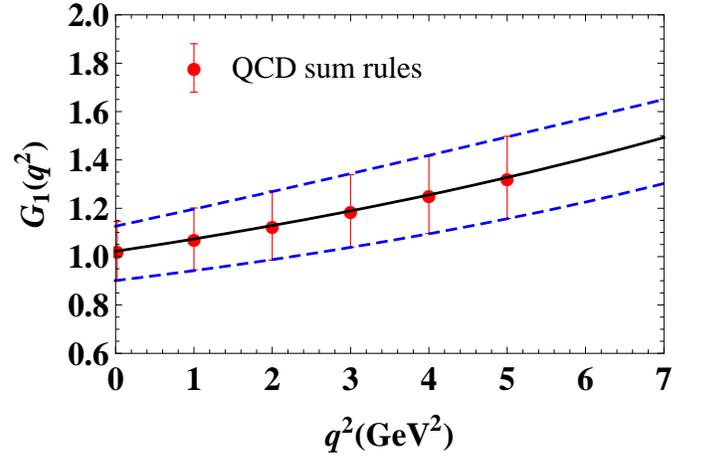}
\caption{Dependence of the weak form factor $G_{1}(q^{2})$ on $q^{2}$: the
QCD sum rule predictions and the fit function $\mathcal{G}_{1}(q^{2})$. The
solid line corresponds to the central values of the parameters $\mathcal{G}%
_{1}^{0},\ g_{1}^{1},\ g_{2}^{1}$, for the upper dashed curve $\mathcal{G}%
_{1}^{0}=1.126,\ g_{1}^{1}=1.792,\ g_{2}^{1}=-0.875$, whereas for the lower
dashed line $\mathcal{G}_{1}^{0}=0.901,\ g_{1}^{1}=1.255,\ g_{2}^{1}=1.106$.
}
\label{fig:G1}
\end{figure}
\begin{figure}[h]
\includegraphics[width=8.8cm]{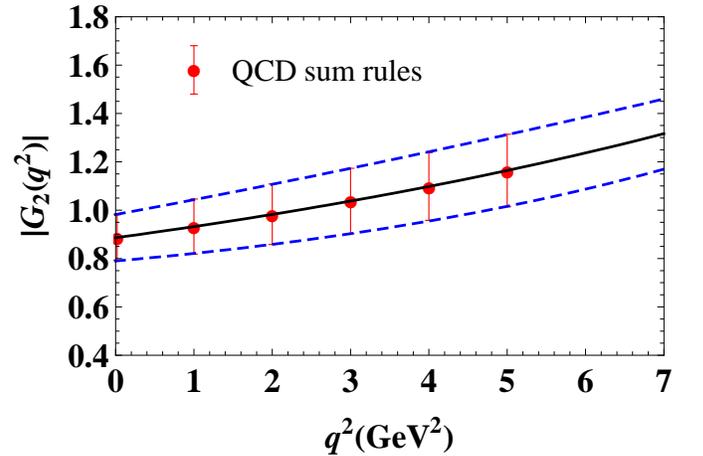}
\caption{The form factor $|G_{2}(q^{2})|$. The solid line describes the
central function. Parameters of the upper and lower dashed curves are $%
\mathcal{G}_{1}^{0}=-0.982,\ g_{1}^{1}=1.771,\ g_{2}^{1}=-0.545$, and $%
\mathcal{G}_{1}^{0}=-0.790,\ g_{1}^{1}=1.039,\ g_{2}^{1}=2.414$,
respectively.}
\label{fig:G2}
\end{figure}

For the fit functions we choose the following analytic expressions
\begin{equation}
\mathcal{G}_{i}(q^{2})=\mathcal{G}_{i}^{0}\exp \left[ g_{1}^{i}\frac{q^{2}}{%
m^{2}}+g_{2}^{i}\left( \frac{q^{2}}{m^{2}}\right) ^{2}\right] .
\label{eq:FFunctions}
\end{equation}%
In Figs.\ \ref{fig:G1} and \ref{fig:G2} one can see the QCD sum rule
predictions for the form factors $G_{1}(q^{2})$ and $|G_{2}(q^{2})|$, in
which ambiguities of computations are shown as error bars. Using the central
values of the form factors and a standard fitting procedure, for the
parameters of the functions $\mathcal{G}_{1}(q^{2})$ and $\mathcal{G}%
_{2}(q^{2})$ we get
\begin{eqnarray}
\mathcal{G}_{1}^{0} &=&1.022,\ g_{1}^{1}=1.383,\ g_{2}^{1}=0.756,  \notag \\
\mathcal{G}_{2}^{0} &=&-0.886,\ g_{1}^{2}=1.440,\ g_{2}^{2}=0.813.
\label{eq:FitPar}
\end{eqnarray}%
The upper and lower limits of the sum rule results are employed to find
corresponding extrapolating functions, plotted in the figures in the form of
dashed curves. Various combinations of these functions are used to estimate
theoretical errors of the semileptonic processes' partial widths.

The differential decay rate $d\Gamma /dq^{2}$ of the process $%
T_{bs}^{-}\rightarrow T_{cs}^{0}l\overline{\nu }_{l}$ can be calculated
using the expression derived in Ref.\ \cite{Sundu:2019feu}, where one needs
to replace parameters of the tetraquarks and weak form factors. Calculations
yield the following predictions
\begin{eqnarray}
\Gamma \left( T_{bs}^{-}\rightarrow T_{cs}^{0}e^{-}\overline{\nu }%
_{e}\right) &=&\left( 1.55\pm 0.42\right) \times 10^{-10}\ \mathrm{MeV},
\notag \\
\Gamma \left( T_{bs}^{-}\rightarrow T_{cs}^{0}\mu ^{-}\overline{\nu }_{\mu
}\right) &=&\left( 1.54\pm 0.42\right) \times 10^{-10}\ \mathrm{MeV},  \notag
\\
\Gamma \left( T_{bs}^{-}\rightarrow T_{cs}^{0}\tau ^{-}\overline{\nu }_{\tau
}\right) &=&\left( 1.91\pm 0.54\right) \times 10^{-11}\ \mathrm{MeV}.  \notag
\\
&&  \label{eq:WeakDecays}
\end{eqnarray}%
Then, for the full width and mean lifetime of the tetraquark $T_{bs}^{-}$ we
find%
\begin{eqnarray}
\Gamma _{\mathrm{full}} &=&\left( 3.28\pm 0.60\right) \times 10^{-10}\
\mathrm{MeV},  \notag \\
\tau &=&2.01_{-0.31}^{+0.44}\times 10^{-12}\ \text{\textrm{s}}.
\label{eq:MLT}
\end{eqnarray}%
Branching ratios of the processes $T_{bs}^{-}\rightarrow D^{0}\overline{K}%
^{0}l\overline{\nu }_{l}$ and $T_{bs}^{-}\rightarrow D^{+}K^{-}l\overline{%
\nu }_{l}$ can be found using $\mathcal{BR}\left( T_{bs}^{-}\rightarrow
T_{cs}^{0}l\overline{\nu }_{l}\right) $ and $\mathcal{BR}\left(
T_{cs}^{0}\rightarrow D^{0}\overline{K}^{0}\right) \simeq \mathcal{BR}\left(
T_{cs}^{0}\rightarrow D^{+}K^{-}\right) \simeq 0.5$. Results of these
computations are collected in Table \ref{tab:BR}.
\begin{table}[tbp]
\begin{tabular}{|c|c|}
\hline\hline
Channels & $\mathcal{BR}$ \\ \hline\hline
$T_{bs}^{-} \to D^{0}\overline{K}^{0}e^{-}\overline{\nu}_e$ & $0.24$ \\
$T_{bs}^{-} \to D^{+}K^{-}e^{-}\overline{\nu}_e$ & $0.24$ \\
$T_{bs}^{-} \to D^{0}\overline{K}^{0}\mu^{-}\overline{\nu}_{\mu}$ & $0.23$
\\
$T_{bs}^{-} \to D^{+}K^{-}\mu^{-}\overline{\nu}_{\mu}$ & $0.23$ \\
$T_{bs}^{-} \to D^{0}\overline{K}^{0}\tau^{-}\overline{\nu}_{\tau}$ & $0.03$
\\
$T_{bs}^{-} \to D^{+}K^{-}\tau^{-}\overline{\nu}_{\tau}$ & $0.03$ \\
\hline\hline
$T_{bb}^{-} \to D^{0}\overline{K}^{0}L$ & $1.7\times 10^{-2}$ \\
$T_{bb}^{-} \to D^{+}K^{-}L$ & $1.6\times 10^{-2}$ \\
$T_{bb}^{-} \to D^{0}\overline{K}^{0}\pi^{+}e^{-}e^{-}$ & $9.8\times 10^{-3}$
\\
$T_{bb}^{-} \to D^{0}\overline{K}^{0}K^{+}e^{-}e^{-}$ & $1.3\times 10^{-3}$
\\
$T_{bb}^{-} \to D^{+}K^{-}\pi^{+}e^{-}e^{-}$ & $9.4\times 10^{-3}$ \\
$T_{bb}^{-} \to D^{+}K^{-}K^{+}e^{-}e^{-}$ & $1.3\times 10^{-3}$ \\
\hline\hline
\end{tabular}%
\caption{The decay channels of the tetraquarks $T_{bs}^{-}$ and $T_{bb}^{-}$%
, and their branching ratios. Above we have used $L=e^{-}e^{+}e^{-}$.}
\label{tab:BR}
\end{table}

%%%%%%%%%%%%%%%%%%%%%%%%%%%%%%%%%%%%%%%%%%%%%%%%%%%%%%%%%%%%%%%%%%

\section{Analysis and conclusions}

\label{sec:AC}
%%%%%%%%%%%%%%%%%%%%%%%%%%%%%%%%%%%%%%%%%%%%%%%%%%%%%%%%%%%%%%
In the present work we have calculated width and mean lifetime of the
tetraquark $T_{bs}^{-}$, which is stable against the strong and
electromagnetic decays. To this end, we have computed partial widths of its
dominant semileptonic decays $T_{bs}^{-}\rightarrow T_{cs}^{0}l\overline{\nu
}_{l}$, where $l$ is one of $e,\ \mu $ and $\tau $ leptons. The tetraquark $%
T_{cs}^{0}$ appeared at the final state of this process is the
strong-interaction unstable particle and decays to conventional mesons $D^{0}%
\overline{K}^{0}$ and $D^{+}K^{-}$. We have also evaluated the spectroscopic
parameters of $T_{cs}^{0}$ and computed the partial widths of its strong
decays, which allowed us to find the branching ratios of the processes $%
T_{bs}^{-}\rightarrow D^{0}\overline{K}^{0}l\overline{\nu }_{l}$ and $%
T_{bs}^{-}\rightarrow D^{+}K^{-}l\overline{\nu }_{l}$. Predictions for the
mass of $T_{bs}^{-}$ obtained in our previous work \cite{Sundu:2019feu}, and
results for the full widths and mean lifetimes of the tetraquarks $%
T_{bs}^{-} $ and $T_{cs}^{0}$ provide a basis for their experimental
investigations.

\begin{figure}[h]
\centering \includegraphics[width=5.0cm]{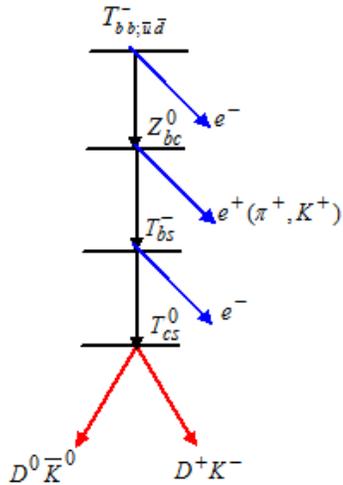}
\caption{Some of the decay modes of the tetraquark $T_{bb;\overline{u}%
\overline{d}}^{-}$.}
\label{fig:Decay}
\end{figure}

But, information gained in the present article is important also to trace
back transformations of the state $T_{bb}^{-}$. Stable nature of the $%
T_{bb}^{-}$ was explored and confirmed by different methods and authors.
This state transforms in accordance with the chains of decays $%
T_{bb}^{-}\rightarrow Z_{bc}^{0}l\bar{\nu _{l}}\rightarrow T_{bs}^{-}l\bar{%
\nu _{l}}\overline{l^{\prime }}\nu _{l^{\prime }}$ and $T_{bb}^{-}%
\rightarrow Z_{bc}^{0}l\bar{\nu _{l}}\rightarrow T_{bs}^{-}Pl\bar{\nu _{l}}$%
, where we take into account both the semileptonic and nonleptonic decays of
the scalar tetraquark $Z_{bc}^{0}$ \cite{Sundu:2019feu}. Now with
information on decays of the tetraquark $T_{bs}^{-}$ at hands, we can fix
some of decay channels of $T_{bb}^{-}$ to conventional mesons. It is not
difficult to see, that $T_{bb}^{-}\rightarrow D^{0}\overline{K}^{0}l\bar{\nu
_{l}}\overline{l^{\prime }}\nu _{l^{\prime }}l^{\prime \prime }\overline{\nu
}_{l^{\prime \prime }}$, $T_{bb}^{-}\rightarrow D^{+}K^{-}l\bar{\nu _{l}}%
\overline{l^{\prime }}\nu _{l^{\prime }}l^{\prime \prime }\overline{\nu }%
_{l^{\prime \prime }}$, $T_{bb}^{-}\rightarrow D^{0}\overline{K}^{0}Pl\bar{%
\nu _{l}}l^{\prime \prime }\overline{\nu }_{l^{\prime \prime }}$, and $%
T_{bb}^{-}\rightarrow D^{+}K^{-}Pl\bar{\nu _{l}}l^{\prime \prime }\overline{%
\nu }_{l^{\prime \prime }}$ are among important modes of such
transformations. In Fig.\ \ref{fig:Decay} we depict some of such channels,
which at the second leg of weak transformations contain products of
semileptonic and nonleptonic decays of $Z_{bc}^{0}$. Their branching ratios
can be found using results of Refs.\ \cite{Agaev:2018khe,Sundu:2019feu} and
information obtained in the present work. For the decay mode $%
T_{bb}^{-}\rightarrow D^{0}\overline{K}^{0}L$ these computations yield%
\begin{eqnarray}
\mathcal{BR}\left( T_{bb}^{-}\rightarrow D^{0}\overline{K}^{0}L\right) &=&%
\mathcal{BR}\left( T_{bb}^{-}\rightarrow T_{bs}^{-}e^{-}e^{+}\right)  \notag
\\
\times \mathcal{BR}\left( T_{bs}^{-}\rightarrow D^{0}\overline{K}%
^{0}e^{-}\right) &=&1.7\times 10^{-2}.  \label{eq:BR1}
\end{eqnarray}%
For simplicity, we have denoted $L=e^{-}e^{+}e^{-}$ and omitted final-state
neutrinos. The branching ratios of other processes [shown in Fig.\ \ref%
{fig:Decay}] are also collected in Table \ref{tab:BR}. The decay channels of
$T_{bb}^{-}$ containing $\mu $ and $\tau $ leptons, and mixed\ modes with $%
e\mu $, $e\tau $, $\mu \tau $, and $e\mu \tau $ leptons at the final state
can be analyzed by the same manner.

The results for the width and lifetime of the tetraquark $T_{bs}^{-}$, and
predictions for branching ratios of $T_{bs}^{-}$ and $T_{bb}^{-}$ have been
obtained using their dominant semileptonic decays. In the case of weak
transformations of the tetraquark $T_{bb}^{-}$ we took into account the
nonleptonic decays of the scalar state $Z_{bc}^{0}$. \ During the present
analysis we have neglected nonleptonic decay modes of $T_{bs}^{-}$ and $%
T_{bb}^{-}$. Our investigations show that branching ratios of nonleptonic
channels are suppressed relative to semileptonic ones \cite%
{Sundu:2019feu,Agaev:2019kkz}, nevertheless, by including into consideration
these modes one can refine the prediction (\ref{eq:BR1}) and ones presented
in Table \ref{tab:BR}.

We also ignored subdominant decay channels which may be generated by weak
decays of heavy quarks, and which are suppressed due to smallness of the
relevant CKM matrix elements. At earlier levels of the weak cascade some of
these modes might create unstable 4-quarks that dissociate to other than $D$
and $K$ mesons.

Finally, in the present work the exotic meson $T_{bs}^{-}$ has been treated
as a scalar particle. But in the decay $Z_{bc}^{0}\rightarrow T_{bs}^{-}%
\overline{l}\nu _{l}$ the final-state tetraquark may bear also other quantum
numbers. By including into analysis these options one may reveal new decay
modes of $Z_{bc}^{0}$, and, hence of $T_{bb}^{-}$. Investigation of these
alternative decays can add a valuable new information on features of the
exotic mesons $T_{bb}^{-}$ and $T_{bs}^{-}$.

\end{document}